\documentclass{article}
\usepackage{graphicx}

\title
{A Path Integral Approach to Derivative Security Pricing:\\
II. Numerical Methods}

\author{Marco Rosa-Clot and Stefano Taddei\\
{\small\em Dipartimento di Fisica, Universit\`a degli Studi di Firenze}\\
{\small\em and Istituto Nazionale di Fisica Nucleare, Sezione di
Firenze,}\\
{\small\em Largo Enrico Fermi 2, I-50125, Firenze, Italy}}

\date{}

\begin{document}

\maketitle

\begin{abstract}

We discuss two numerical methods, based on a path integral
approach described in a previous paper (I), for solving the stochastic
equations underlying the financial markets:
the Monte Carlo
approach, and the Green function deterministic numerical method.
Then, we apply the latter to some specific financial
problems. In particular, we consider the pricing of a European option,
a zero-coupon bond, a caplet, an American option, and a Bermudan swaption.

\end{abstract}

\section{Introduction}

The evolution law of a financial index, $X$, is often given by a
stochastic differential equation, which can be discretized as
\begin{equation} 
\Delta X=A(X(t),t)\;\Delta t+ 
\sigma(X(t),t)\;\Delta W, 
\label{e:1.3} 
\end{equation}
where $\Delta t$ is the time step, and $\Delta W$
is a Wiener process increment.
(In this paper, the random variables are denoted by capital letters,
and the ordinary variables by small ones.
Moreover, for the sake of simplicity, we consider only the one-dimensional
case, but the extension
to the multi-dimensional case is straightforward.)
In a previous paper\cite{rct1}, hereafter referred as paper I,
we have seen that the continuous limit of equation (\ref{e:1.3})
cannot be written in unambiguous form, i.e. it
is well defined only if also a discretization rule is given. Therefore,
from now on, we will always write the underlying stochastic equation in
the discretized form (\ref{e:1.3}), understanding that the continuous
limit must be taken.

Many financial quantities can be defined as the conditional
expectation value of some
functional, $g[X(\tau)]$, of the stochastic process, $X(\tau)$,
obeying the equation (\ref{e:1.3}).
This conditional expectation value can be written as (see, paper I)
\begin{eqnarray}
\lefteqn{\!\!\!\!E[g[X(\tau)]\mid X(t_0)=x_0]=}\nonumber\\
&&\!\!\!\!\!\!\!\!\!\!\!\!\!\!\int_{-\infty}^{+\infty}\!\!\! dx_N
{\int\!\!\!\int}^{x(t_N)=x_N}_{x(t_0)=x_0}\!
{\cal D}[\sigma(x,\tau)^{-1}x(\tau)]\;g[x(\tau)]
\exp\!\left\{\!-\!\int_{t_0}^{t_N}\!\!\! L[x(\tau),{\dot
x}(\tau);\tau]d\tau\right\}\!.
\label{e:path}
\end{eqnarray}
The R.H.S. of Eq. (\ref{e:path}) is called path integral\cite{feynman-hibbs},
and the functional measure, ${\cal D}[\sigma(x,\tau)^{-1}x(\tau)]$, means
summation on all possible random paths from
$X(t_0) = x_0$ to $X(t_N)=x_N$, where $t_0$, and $t_N$ are the initial and
final times, respectively. In general,
the Lagrangian function, $L[x,{\dot x};\tau]$, is not defined univocally.
A possible expression is given by
\begin{equation}
L[x(\tau),{\dot x}(\tau);\tau]=
{1\over 2\;\sigma(x,\tau)^2}\;
[{\dot x}-A(x,\tau)]^2.
\label{e:lagra}
\end{equation}
The analytical methods discussed in paper I do not go behind the quadratic
Lagrangians.
This drawback can be partially overcome by using approximate analytical
techniques
such as the perturbative expansion or the saddle point approximation, but
a more practical and general approach is the numerical one.

The aim of this paper is to describe two numerical techniques to evaluate
the path integral above: (a) the Monte Carlo method, which is very
general and powerful, but has a low precision, and high CPU time requirements;
(b) the Green function deterministic numerical method (GFDNM),
which has been recently 
developed\cite{rct2,rct2b,taddei,rct3}, and it is an advantageous alternative
when the stochastic process has a low dimensionality.

In section \ref{s:2}, we recall some general notions on probability theory,
and we show that the introduction of the conditional expectation
value (\ref{e:path}) as a path integral is just a generalization of the usual
concept. In sections \ref{s:3}, and \ref{s:4}, we describe the Monte Carlo,
and the Green function deterministic numerical methods, respectively.
Finally, in section \ref{s:5},
some applications of the latter method are discussed.

\section{Preliminary notions on probability theory}
\label{s:2}

A stochastic process can be defined as a random variable, $X(\tau)$, which is
function of either a discrete or a continuous {\em time} variable, $\tau$.
The statistical properties of the random variable, $X(t)$,
at a fixed time, $t$, are
determined by the probability density function, or probability distribution
function, $\rho(x,t)$. The stochastic process as a whole is characterized by a
set of joint probability density functions, eventually infinite,
\begin{eqnarray}
&&\rho(x_0,t_0)\nonumber\\
&&\rho(x_1,t_1;x_0,t_0)\nonumber\\
&&\rho(x_2,t_2;x_1,t_1;x_0,t_0)\label{e:pt1}\\
&&\ldots,\nonumber
\end{eqnarray}
with $t_i\in[t_0,t_N]$, which must satisfy the Kolmogorov compatibility conditions,
\begin{eqnarray}
\lefteqn{\int\rho(x_N,t_N;\ldots;x_i,t_i;\ldots;x_0,t_0)\;dx_i=}\nonumber\\
&&\;\;\;\;\rho(x_N,t_N;\dots;x_{i+1},t_{i+1};x_{i-1},t_{i-1};\dots;x_0,t_0)\;.
\label{e:pt1bis} 
\end{eqnarray}
The conditional probability density functions are defined as the
ratios
\begin{equation}
\rho(x_N,t_N;\ldots;x_i,t_i\mid x_{i-1},t_{i-1};\ldots;x_0,t_0)=
{\rho(x_N,t_N;\ldots;x_0,t_0)\over \rho(x_{i-1},t_{i-1};\ldots;x_0,t_0)}\;.
\label{e:pt2}
\end{equation}

Let us now assume that the time variable is discrete.
We can define the expectation value of a function of the stochastic process,
$g(X(t_N),\ldots,X(t_0))$, as
\begin{eqnarray}
\lefteqn{\!\!\!\!\!\!\!\!\!\!\!\!\!\!E[g(X(t_N),\ldots,X(t_0))]=}\nonumber\\
&&\int\!\!\dots\!\!\int dx_N\ldots dx_0\;g(x_N,\ldots,x_0)\;
\rho(x_N,t_N;\ldots;x_0,t_0)\;,
\label{e:pt3}
\end{eqnarray} 
and the conditional expectation value as
\begin{eqnarray}
\lefteqn{\!\!\!\!\!\!\!E[g(X(t_N),\ldots,X(t_0))\mid
X(t_{i-1})=x_{i-1};\ldots;X(t_0)=x_0]=}\nonumber\\
&&\!\!\!\!\!\!\!\!\!\!\!\!\!\!\!\!\!\int\!\!\dots\!\!\int dx_N\ldots dx_i\;
g(x_N,\ldots,x_0)\;
\rho(x_N,t_N;\ldots;x_i,t_i\mid x_{i-1},t_{i-1};\ldots;x_0,t_0).
\label{e:pt4}
\end{eqnarray}
The equation above defines a function of the ordinary variables,
$x_{i-1},\ldots,x_0$.
If these variables are substituted by the random variables,
$X(t_{i-1}),\ldots,X(t_0)$, the conditional expectation value
(\ref{e:pt4}) becomes a function of the stochastic process.
In this case we will use the shorter notation
\begin{eqnarray}
\lefteqn{\!\!\!\!\!\!\!E[g(X(t_N),\ldots,X(t_0))\mid
X(t_{i-1});\ldots;X(t_0)\,]\equiv}\nonumber\\
&&E[g(X(t_N),\ldots,X(t_0))\mid
X(t_{i-1})=X(t_{i-1});\ldots;X(t_0)=X(t_0)].
\end{eqnarray}
Finally, we can define the expectation value with fixed initial and
final points
\begin{eqnarray}
\lefteqn{\!\!\!\!\!\!\langle x_N,t_N\mid g(X(t_N),\ldots,X(t_0))
\mid x_0,t_0\rangle=}\nonumber\\
&&\int\!\!\dots\!\!\int dx_{N-1}\ldots dx_1\;g(x_N,\ldots,x_0)\;
\rho(x_N,t_N;\ldots;x_1,t_1\mid x_0,t_0).
\label{e:pt5}
\end{eqnarray} 
Some relations among the quantities above are
\begin{eqnarray}
\lefteqn{E[g(X(t_N),\ldots,X(t_0))]=}\nonumber\\
&&\!\!\!\!\!\!\!\!\!\!\!\int\!\!\dots\!\!\int dx_{i-1}\ldots dx_0\;
E[g(X(t_N),\ldots,X(t_0))\mid
X(t_{i-1})=x_{i-1};\ldots;X(t_0)=x_0]\nonumber\\
&&\;\;\;\;\;\;\;\;\;\;\;\;\;\;\;\;\;\;\;\;\;\;\;\;\;\;\;\;\;\;\;\;
\;\;\;\;\;\;\;\;\;\;\;\;\;\;\;\;\;\;\;\;\;\times\;\;
\rho(x_{i-1},t_{i-1};\ldots;x_0,t_0)\;,
\label{e:pt6}
\end{eqnarray}
and
\begin{eqnarray}
\lefteqn{\!\!\!\!\!\!\!\!\!\!\!\!\!\!\!\!\!\!\!\!
E[g(X(t_N),\ldots,X(t_0))\mid X(t_0)=x_0]=}\nonumber\\
&&\;\;\;\;\;\;\;\;\;\;\;\;\;\;\;\;\;\;\;\;\;\;
\int dx_N\;\langle x_N,t_N\mid g(x_N,\ldots,x_0)\mid x_0,t_0\rangle\;.
\label{e:pt7}
\end{eqnarray} 

Let us now consider two particular types of
stochastic processes: the Markov process, and the Gaussian process.

The Markov process is defined as
a stochastic process such that
\begin{equation}
\rho(x_i,t_i\mid x_{i-1},t_{i-1};\ldots;x_0,t_0)=
\rho(x_i,t_i\mid x_{i-1},t_{i-1}),
\label{e:pt2mar}
\end{equation}
i.e. the conditional probability density function at the time, $t_i$, depends
only on the next earlier time, and not on the whole previous history of the
process. It then follows that the conditional probability density with
fixed initial point can be written as
\begin{equation}
\rho(x_N,t_N;\ldots;x_1,t_1\mid x_0,t_0)=
\prod_{i=1}^{N} \rho(x_i,t_i\mid x_{i-1},t_{i-1}).
\label{e:markov}
\end{equation}
The function, $\rho(x_i,t_i\mid x_{i-1},t_{i-1})$, is also called
transition probability.
The following properties of a Markov process can be easily proved
\begin{eqnarray}
\lefteqn{\!\!\!\!\!\!\!\!\!\!\!\!\!\!
E[E[g_1(X(t_N),\ldots,X(t_i))\mid X(t_i)\,]
\;g_2(X(t_i),\ldots,X(t_0))\mid X(t_0)=x_0]=}\nonumber\\
&&\;\;\;\;\;E[g_1(X(t_N),\ldots,X(t_i))\;g_2(X(t_i),\ldots,X(t_0))
\mid X(t_0)=x_0],
\label{e:inner}
\end{eqnarray}
and
\begin{eqnarray}
\lefteqn{\!\!\!\!\!\!\!\!\!\!\!\!\!\!
E[g(X(t_N),\ldots,X(t_i))\mid X(t_i)=x_i\,;\;X(t)=x]=}\nonumber\\
&&\;\;\;\;\;\;\;\;\;\;\;\;\;\;\;\;\;\;\;\;\;\;\;\;\;\;\;\;\;\;
E[g(X(t_N),\ldots,X(t_i))\mid X(t_i)=x_i],
\label{e:final}
\end{eqnarray}
for every $t<t_i$.

Finally, a stochastic process is called Gaussian process if all its joint
probability density functions are multivariate Gaussian distributions.

\subsection{Expectation value of a functional of random paths}

If the time variable, $\tau$, is continuous, instead of a function of a
stochastic process we should more properly speak of a functional, $g[X(\tau)]$
(note the square brackets in the notation), of a random path.
Its expectation value and its conditional expectation values can be written
as a generalization of the equations in the previous section, i.e.
\begin{equation}
E[g[X(\tau)]]=\lim_{N\rightarrow\infty}
\int\!\!\dots\!\!\int \prod_{i=0}^N dx_i\;g(x_N,\ldots,x_0)\;
\rho(x_N,t_N;\ldots;x_0,t_0),
\label{e:dis1}
\end{equation}

\begin{eqnarray}
\lefteqn{\!\!\!\!\!\!
\langle x_N,t_N\mid g[x(\tau)]\mid x_0,t_0\rangle}\nonumber\\
&&=\lim_{N\rightarrow\infty}
\int\!\!\dots\!\!\int \prod_{i=1}^{N-1} dx_i\;g(x_N,\ldots,x_0)\;
\rho(x_N,t_N;\ldots;x_1,t_1\mid x_0,t_0)\label{e:dis2}\\
&&\equiv\int\!\!\!\int_{x(t_0)=x_0}^{x(t_N)=x_N}
{\cal D}[x(\tau)]\;g[x(\tau)]\;\rho[x(\tau)],
\label{e:con2}
\end{eqnarray}
and
\begin{equation}
E[g[X(\tau)]\mid X(t_0)=x_0]=\int_{-\infty}^{+\infty} dx_N
\langle x_N,t_N\mid g[x(\tau)]\mid x_0,t_0\rangle\;.
\label{e:relat}
\end{equation}
Moreover, for a Markov process, we have a straightforward generalization
of Eqs. (\ref{e:inner}), and (\ref{e:final}), i.e.
\begin{eqnarray}
\lefteqn{\!\!\!\!\!\!\!\!\!\!\!\!
E\Bigl[E\Bigl[g_1[X(\tau)\mid\tau\in[t_i,t_N]\,]\mid X(t_i)\,\Bigl]
\;g_2[X(\tau)\mid\tau\in[t_0,t_i]\,]\mid X(t_0)=x_0\Bigl]\,=}\nonumber\\
&&\;\;\;\;\;E\Bigl[g_1[X(\tau)\mid\tau\in[t_i,t_N]\,]\;
g_2[X(\tau)\mid\tau\in[t_0,t_i]\,]\mid X(t_0)=x_0\Bigl],
\label{e:innerpath}
\end{eqnarray}
and
\begin{eqnarray}
\lefteqn{\!\!\!\!\!\!\!\!\!\!\!\!
E\Bigl[g[X(\tau)\mid\tau\in[t_i,t_N]\,]\mid  X(t_i)=x_i\,;\; X(t)=x\Bigl]\;=}
\nonumber\\
&&\;\;\;\;\;\;\;\;\;\;\;\;\;\;\;\;\;\;\;\;\;\;\;\;\;\;\;\;\;
E\Bigl[g[X(\tau)\mid\tau\in[t_i,t_N]\,]\mid  X(t_i)=x_i],
\label{e:finalpath}
\end{eqnarray}
where $t<t_i$, and, for the sake of clarity, we have explicitly written the
intervals of variation of the time variable, $\tau$.

The expression (\ref{e:dis2}) represents a definition of the path integral
(\ref{e:con2}), but the latter is only a formal way
to write the limit of the former discretized expression.
However, for a Wiener process, and more generally,
for a Langevin process, described in the next section,
an exact mathematical measure can be defined (Wiener integral)\cite{wiener}.

The functions $g(x_N,\ldots,x_0)$,
and $\rho(x_N,t_N;\ldots;x_1,t_1\mid x_0,t_0)$
represent a discretization
of the functionals, $g[x(\tau)]$, and $\rho[x(\tau)]$. We want to stress here
that, in general, both
the explicit form of the functionals and the discretization
procedure are not defined univocally. A more detailed discussion of this
problem can be found in paper I.

\subsection{Wiener and Langevin processes}

A stochastic process can be Gaussian and Markovian at the same time.
The Wiener process, for example, is defined by the initial
probability density function, $\rho(x,t_0)=\delta(x)$, and the Gaussian
transition probability,
\begin{equation}
\rho(x_i,t_i\mid x_{i-1},t_{i-1})=
{1\over\sqrt{2\pi\,\sigma^2\,\Delta t}}\;\exp\left\{-{(x_i-x_{i-1})^2
\over 2\;\sigma^2\,\Delta t}\right\},
\label{e:brown}
\end{equation}
where $\Delta t=t_i-t_{i-1}$.

Another Markov, in general non-Gaussian, process is the solution
of the Langevin equation (\ref{e:1.3}).
An explicit expression for
the joint probability density function corresponding to
this equation does not always exist.
However, a general expression of the transition
probability for small time steps (short-time transition probability),
correct up to $O(\Delta t)$, is given by (see, paper I)
\begin{eqnarray}
\lefteqn{\!\!\!\!\!\!\!\!\rho(x_i,t_i\mid x_{i-1},t_{i-1})\simeq}\nonumber\\
&&{1\over\sqrt{2\pi\,\sigma(x_{i-1},t_{i-1})^2\,\Delta t}}\;
\exp\left\{-{(x_i-x_{i-1}-A(x_{i-1},t_{i-1})\;\Delta t)^2
\over 2\;\sigma(x_{i-1},t_{i-1})^2\;\Delta t}\right\}.
\label{e:lang}
\end{eqnarray}
Therefore,
\begin{eqnarray}
\lefteqn{\!\!\!\!\!\!\!\!\!\!\!\!\!\!\!\!\!\!\!\!
\rho(x_N,t_N;\ldots;x_1,t_1\mid x_0,t_0)
\simeq(2\pi\,\sigma(x_{i-1},t_{i-1})^2\,\Delta t)^{-N/2}}\nonumber\\
&&\;\;\;\;\;\exp\left\{-\sum_{i=1}^N {(x_i-x_{i-1}-
A(x_{i-1},t_{i-1})\,\Delta t)^2
\over 2\;\sigma(x_{i-1},t_{i-1})^2\,\Delta t}\right\}.
\label{e:distr}
\end{eqnarray}
Note that the expression (\ref{e:lang}) is equal to
(\ref{e:brown}) when $A=0$, and $\sigma={\rm const}$.
In this case the short-time transition probability  (\ref{e:lang})
is exact, and the Wiener process is the solution of the Langevin equation.

From now on, if not explicitly specified,
we will consider only Langevin processes.
The probability density functional for
a Langevin process, apart from a normalization factor usually
included into the measure, can be written as
\begin{equation}
\rho[x(\tau)]\sim
\exp\left\{-\int_{t_0}^{t_N}L[x(\tau),{\dot x}(\tau);\tau]
d\tau\right\};
\label{e:es}
\end{equation}
then the Eq. (\ref{e:relat}) becomes equal to the
conditional expectation value (\ref{e:path}).
We recall that the Lagrangian (\ref{e:lagra}), with the
discretization rule (\ref{e:distr}), corresponds to the pre-point formulation
of the path integral (see, paper I).

\section{The  Monte Carlo method}
\label{s:3}

Monte Carlo is a technique for the numerical computation of mathematical
quantities using {\em random numbers} \cite{ulam}. Its basic idea lies on two
important limit theorems of probability theory: the
{\em law of large numbers}, and the
{\em central limit theorem}\/. As a result of these theorems, if
$f(X)$ is a function of a random variable, $X$, and
the random numbers, $x^{(r)}$, are sampled from its probability distribution,
for large $M$ the average
\begin{equation}
m={1\over M}\;\sum_{r=1}^M f(x^{(r)}),
\label{e:pt8}
\end{equation}
is an estimator of $\mu=E[f(X)]$.
Moreover, even if the random variable has a quite general probability
distribution, the averages obtained by different samplings
are distributed according to a normal probability density function with
mean $\mu$, and variance
$\displaystyle{\sigma_{\mu}^2=E\left[{(f(X)-\mu)^2\over M}\right]}$.
Finally, an estimation of $\sigma_{\mu}^2$ is given by
the quantity,
\begin{equation}
\displaystyle{s_m^2={1\over M(M-1)}\sum_{r=1}^M (f(x^{(r)})-m)^2}.
\end{equation}

These theorems can be generalized to the case of multivariate
probability distribution functions.

\subsection{Monte Carlo integration}

The aim of this introductory section is to describe the use of the Monte
Carlo method to perform the numerical integration of an ordinary
function. We briefly discuss both the plain and the importance sampling
methods.

\subsubsection{Plain integration}

Let us consider the ordinary integral,
\begin{equation}
I=\int_a^bf(x)\:dx.
\label{e:1}
\end{equation}
If $x^{(1)},\ldots,x^{(M)}$ are $M$ random numbers uniformly distributed in the
interval $[a,b]$, as a result of the limit theorems, we have
\begin{equation}
I=E_u[f(X)]\simeq\overline{I}={1\over M}\sum_{r=1}^Mf(x^{(r)})
\end{equation}
(the indices, $u$, means that the expectation value is computed on the uniform
probability distribution), and the statistical error is
of the order of the standard deviation,
\begin{equation}
\sigma_u\simeq\sqrt{{1\over M(M-1)}\sum_{r=1}^M (f(x^{(r)})-\overline{I})^2}.
\label{e:err}
\end{equation}

\subsubsection{Importance sampling}

We must often calculate an integral of the form,
\begin{equation}
I=\int_a^bf(x)\,\rho(x)\:dx,
\label{e:3}
\end{equation}
where $\rho(x)$ is some probability density function defined in the interval
$[a,b]$. It can happen that in the most part of the interval, the function,
$\rho(x)$, is almost zero. In this case, with the uniform sampling defined
above, a large number of points gives a negligible contribution.
In order to overcome this limit, we can use the
importance sampling method: if $x^{(1)},\ldots,x^{(M)}$ are random
numbers sampled from the distribution function, $\rho(x)$, for the
limit theorems again, we have
\begin{equation}
I=E_\rho[f(X)]\simeq
\tilde{I}={1\over M}\sum_{r=1}^Mf(x^{(r)}),
\end{equation}
and the statistical error,
\begin{equation}
\sigma_\rho\simeq
\sqrt{{1\over M(M-1)}\sum_{r=1}^M (f(x^{(r)})-\tilde{I})^2}.
\label{e:err2}
\end{equation}

\subsection{Random number generation}

The previous section has shown that, in general, we need random numbers sampled
from an arbitrary probability distribution. Uniform random number
generators are implemented on almost all computers, but the sampling
from more complex probability distributions must be performed by
appropriate algorithms. In this section we describe two fundamental algorithms
of this kind.

\subsubsection{Acceptance-rejection algorithm}

This method was proposed by von Neumann\cite{neumann} to obtain random
numbers with a given probability
distribution, $\rho(x)$. Let $\rho(x)$ be defined
in a finite interval $[a,b]$, and bounded by a value, $U$. Then random
numbers with probability distribution, $\rho(x)$, can be obtained by the
following algorithm
\begin{itemize}
\item[i.] Draw $x$ from a uniform distribution in $[a,b]$.

\item[ii.] Draw $p$ from a uniform distribution in $[0,U]$.

\item[iii.] If $p\leq \rho(x)$ then accept x else reject it, and goto i.
\end{itemize}
In order to have a better efficiency, the upper bound, $U$, must be as
close as possible to the maximum of $\rho(x)$. Unfortunately
many points must be rejected, and the method is not very efficient,
especially
in the case of multivariate probability distributions. Moreover, it can be
applied only to bounded distributions with a finite range.

\subsubsection{Metropolis algorithm} 

A more efficient and general method is given by the Metropolis
algorithm\cite{metropolis}.
This is related to an important property of the Markov processes:
after a number of time steps large enough, the final probability density of
a stochastic system, which evolves according to a discrete time Markov process,
is independent on the number of steps, and on the initial configuration.
The Metropolis algorithm is a solution of the inverse problem:
is it possible to
construct a Markov process which, after a number of steps large enough,
yields a configuration with a given probability density function,
$\rho(x)$? The
solution is not unique, but the Metropolis procedure is particularly simple
to implement. It consists of the following steps:
\begin{itemize}
\item[i.] Start at the initial time from a point $x$.

\item[ii.] Draw a random number, $q$, from the uniform distribution in the
interval $[0,1]$, and compute $x^\prime=x+D(2q-1)$, where $D$ is some
arbitrary parameter.

\item[iii.] If $\rho(x^\prime)>\rho(x)$ then goto v.

\item[iv.] Draw another random number, $p$, from the uniform distribution
in the interval $[0,1]$.
If $\rho(x^\prime)\le p\,\rho(x)$ then put $x^\prime=x$.

\item[v.] Put $x^\prime$ in the list of random numbers,
rename $x^\prime$ to $x$,
and goto ii.
\end{itemize}
The random numbers, $x^\prime$, have the required probability density function.
Actually, a certain number of steps must be performed until convergence is
attained, i.e. a certain number of initial random points must be rejected.
The parameter, $D$, is arbitrary, but it has been empirically proved that an
appropriate value of $D$ should
give a ratio of acceptances between 50\% and 70\%.
If we have a multivariate distribution, and hence a multi-dimensional array
of points, each point should be moved in turn.
The new array corresponds to one step of the Markov process.

\subsection{Expectation value computation}

A continuous stochastic process, $X(\tau)$,
can be specified either by its
evolution law (stochastic differential equation), or by
its probability density functional
(path integral formulation). Thus,
it can be approximated either by the
solution of the discretized Langevin equation (\ref{e:1.3}),
or by the discretized stochastic process with the multivariate
probability density function (\ref{e:distr}).

According to the limit theorems, we can write the following
approximate expression for the conditional expectation value (\ref{e:relat}),
\begin{equation}
E[g[X(\tau)]\mid X(t_0)=x_0]\simeq\lim_{M\rightarrow\infty}
\sum_{r=1}^M {g(x_N^{(r)},\ldots,x_0^{(r)})\over M},
\label{e:exval}
\end{equation}
where the random paths, $x_0^{(r)},\ldots,x_N^{(r)}$,
are selected either by solving
Eq. (\ref{e:1.3}), or by finding the stochastic process with the
probability density (\ref{e:distr}); in both cases with a fixed
starting point, $x_0^{(r)}=x_0$.
We will now briefly describe how these two problems can be solved
by the Monte Carlo methods.

\subsubsection{Langevin equation approach}

A solution, $x_0^{(r)},\ldots,x_N^{(r)}$, of the Langevin
equation (\ref{e:1.3}) can be obtained by the following algorithm:
\begin{itemize}
\item[i.] Put $i=0$, and $x^{(r)}_0=x_0$.

\item[ii.] Draw a random number, $z$, from the normal distribution with
mean 0 and variance $\Delta t$.

\item[iii.] Take $x_{i+1}^{(r)} =x_i^{(r)} + A(x_i^{(r)},t_i)\; \Delta t  +
\sigma (x_i^{(r)},t_i)\; z $.

\item[iv.] Store $x_{i+1}^{(r)}$.

\item[v.] If $i<(N-1)$ then put $i=i+1$, and goto ii.
\end{itemize}
An approximation of
the conditional expectation value (\ref{e:relat}) is given by the summation
in Eq. (\ref{e:exval}), where the sum extends to all paths generated.

\subsubsection{Path integral approach}

The discretized conditional expectation value, defined by Eqs.
(\ref{e:dis2}), and (\ref{e:relat}),
is simply a multi-dimensional integral. Therefore, we can use the Metropolis
algorithm to find an ensemble of discretized paths with the multivariate
probability density function (\ref{e:distr})\cite{crfr}. An approximation of
the conditional expectation value (\ref{e:relat}) is given again
by the sum in Eq. (\ref{e:exval}) (see also Ref. \cite{makivic}).

We want to point out here that the Markov process underlying
the Metro\-polis algorithm
must not be confused with the Markov process
describing the {\em physical}\/ process, and obeying the Langevin equation
(\ref{e:1.3}). The former is only a formal device to
obtain the required joint probability density.
A sample path corresponding to the first stochastic process is given by
the set of vectors,
$(x_0^{(1)},\ldots,x_N^{(1)})\,,\ldots, (x_0^{(M)},\ldots,x_N^{(M)})$,
while a sample path for the second process is simply given
by the single array, $x_0^{(r)},\ldots,x_N^{(r)}$.

\section{Green function deterministic numerical me\-thod}
\label{s:4}

The main drawback of the Monte Carlo method is the CPU time requirement.
We must remind that to gain one order of magnitude in
the precision we need to increase the CPU time of two orders.
In this section we describe an alternative method
\cite{rct1,rct2} which has some advantages in low dimensional problems.
In the following we will call indifferently Green function or
transition probability the conditional probability density (\ref{e:pt2mar}),
since actually this function represents the Green function of the partial
derivative equation corresponding to the Langevin equation (\ref{e:1.3}).

\subsection{Expectation value computation}
\label{s:4gen}

Let us consider a general Markov process.
The probability density functional, $\rho[x(\tau)]$,
can be approximated by the product of
transition probabilities given in Eq. (\ref{e:markov}).
If we assume that the functional, $g[x(\tau)]$,
can be discretized in the form,
\begin{equation}
g[x(\tau)]\simeq \prod_{i=0}^{N} g^{(i)}(x_i),
\end{equation}
which includes most of the interesting cases, then an approximation of
the conditional expectation value (\ref{e:relat}) is given by
\begin{equation}
E[g[X(\tau)]\mid X(t_0)=x_0]\simeq \int\!...\!
\int \prod_{i=1}^{N} dx_i\;
g^{(N)}(x_{N})\; \prod_{j=1}^{N} {\tilde\rho}(x_j,t_j\mid x_{j-1},t_{j-1}),
\label{e:condapp}
\end{equation}
where the function, ${\tilde\rho}$, is
\begin{equation}
{\tilde\rho}(x_i,t_i\mid x_{i-1},t_{i-1}) =
\rho(x_i,t_i\mid x_{i-1},t_{i-1})\; g^{(i-1)}(x_{i-1}).
\end{equation}
Let us now consider a single integration
\begin{equation}
\int dx_i\; {\tilde\rho}(x_{i+1},t_{i+1}\mid x_i,t_i)\;
{\tilde\rho}(x_i,t_i\mid x_{i-1},t_{i-1})\;.
\label{e:single}
\end{equation}
If we approximate this integral by using a numerical quadrature
rule, we obtain the following algebraic relation
\begin{equation}
\sum_{\gamma=1}^M\;
{\tilde\rho}^{~(i)}_{\alpha\gamma}\;\;{\tilde\rho}^{~(i-1)}_{\gamma\beta}\;
w_\gamma\;,
\end{equation}
where the matrices, ${\tilde\rho}^{~(i)}$, are defined by
\begin{equation}
{\tilde\rho}^{~(i)}_{\alpha\beta}=
{\tilde\rho}(z_\alpha,t_{i+1}\mid z_\beta,t_i),
\end{equation}
the quantities, $w_\alpha$, and $z_\alpha$, are the weights and the grid
points, respectively, associated with the integration rule, and
$\alpha,\beta,\gamma=1,\ldots,M$.
In conclusion, the expression (\ref{e:condapp}) can be written as
\begin{equation}
E[g[X(\tau)]\mid X(t_0)=z_\alpha]
\simeq \sum_{\gamma_N,\ldots,\gamma_1=1}^M\;g^{(N)}_{\gamma_N}\;\;
G^{(N-1)}_{\gamma_N\gamma_{N-1}}\;\;
G^{(N-2)}_{\gamma_{N-1}\gamma_{N-2}}
\dots G^{(0)}_{\gamma_1\alpha}\;,
\label{e:prodg}
\end{equation}
where $G^{(i)}_{\alpha\beta}=
w_{\alpha}\,{\tilde\rho}^{(i)}_{\alpha\beta}$, and $g^{(N)}_{\gamma_N}=
g^{(N)}(z_{\gamma_N})$.
Therefore, we have reduced the evaluation of the expectation
value of a functional to the product of $N$ matrices with dimension $M$.
By starting the calculation from the
left, we need to memorize just linear arrays,
while the matrix elements, $G^{(i)}_{\alpha\beta}$, can be computed step by
step. In practice, we adopt the following algorithm:
\begin{itemize}
\item[i.] Put $u_\alpha=g^{(N)}_\alpha$ ($\alpha=1,\ldots,M$), and $i=N-1$.

\item[ii.] Put
$v_\alpha=\displaystyle{\sum_{\beta=1}^M}\; u_\beta\,G^{(i)}_{\beta\alpha}$
($\alpha=1,\ldots,M$).

\item[iii.] If $i>0$ then put $u_\alpha=v_\alpha$ ($\alpha=1,\ldots,M$),
$i=i-1$, and goto ii.
\end{itemize}
Here the arrays, $u_\alpha$, and $v_\alpha$, are two working vectors.

\subsubsection{Path dependent options}
\label{s:4pathd}

In sections \ref{s:5.4}, and \ref{s:5.5} we will discuss two examples
of path dependent options.
The pricing of this type of derivative securities requires to evaluate
the expectation value of functionals containing some constraints.
In general, their explicit expression is very
involved, and an exact analytical treatment is not possible. Instead
a pure numerical approach implies only a small difference in the
algorithm described above.
In particular, for the cases considered in this paper, it becomes simply
\begin{itemize}
\item[i.] Put $u_\alpha=g^{(N)}_\alpha$ ($\alpha=1,\ldots,M$), and $i=N-1$.

\item[ii.] Put $w^{(i)}_\alpha=f(z_\alpha,t_i)$\/, and
$v_\alpha=
\max\left(\displaystyle{\;\sum_{\beta=1}^M}\;
u_\beta\,G^{(i)}_{\beta\alpha}\;,\;\;
w^{(i)}_\alpha\;\right)$

($\alpha=1,\ldots,M$).

\item[iii.] If $i>0$ then put $u_\alpha=v_\alpha$ ($\alpha=1,\ldots,M$),
$i=i-1$, and goto ii.
\end{itemize}
The only difference is in the point ii, where we have now a test operation,
and the function, $f(z_\alpha,t_i)$, depends on the problem considered.

\subsection{Computational details}

The method described above is quite general. Let us now consider
the particular case of a stochastic process obeying the Langevin equation
(\ref{e:1.3}). Here we can use the approximate expression (\ref{e:lang})
for the short-time transition probability.

Since the interval of integration in Eq.\ (\ref{e:single})
is infinite, the numerical integration
should be performed by a quadrature rule for
improper integrals (for instance, the Gaussian quadrature).
On the other hand, since the integrand is essentially a narrow
Gaussian of width $\sigma\,\sqrt{\Delta t}$ whose central position, $x_i$,
moves on the whole interval, we need a
grid dense enough to give an accurate quadrature everywhere.
Therefore we are forced to take a uniform distribution of grid points
with a lattice spacing,
$\Delta x = x_i - x_{i-1}\sim{\overline\sigma}\,\sqrt{\Delta t}$; where
${\overline\sigma}=\displaystyle{\min_{x\in{\cal I},\;\tau\in[t_0,t_N]}}
\,(\sigma(x,\tau))$, and
the interval, ${\cal I}$, is the finite range of integration due to
the finite number of points.
As a result the stochastic process is confined in a box, but
an interval large enough
gives negligible corrections. A good choice for the
quadrature formula is the trapezoidal rule, which yields very accurate results
with Gaussian functions, and, in general, with functions which are zero,
with all their derivatives, out of some range.

Actually, the relation between
$\Delta t$ and $\Delta x$, and the need of a finite range of
integration are two connected problems: if we fix the number of
grid points and the interval, ${\cal I}$, then the value of $\Delta t$
is fixed by the relation $\Delta t\sim \Delta x^2/{\overline\sigma}^2$ (in practice,
a ratio ${\overline\sigma}^2\,\Delta t/\Delta x^2=1$ gives an accuracy greater than
$1\%$, and already with 1.25 we get at least 8 digits).
In other words, if we take $\Delta t$ going to zero, the Gaussian
part of the transition probability becomes strongly peaked, and we need
a very large number of points to obtain a good precision.
This means that we cannot take $\Delta t$ as small as possible, and the
systematic error which depends on $\Delta t$ can be significant
(this systematic error must not be confused with the numerical error
in the quadrature, which depends only on the ratio
${\overline\sigma}^2\,\Delta t/\Delta x^2$).

As a final remark, we note that the definition above for $\overline\sigma$
can be
sometimes source of troubles. In particular, when the function, $\sigma(x,t)$,
has some zeros or a very wide range of variation, the time step, $\Delta t$,
can be very large. In this case a larger number of points than usual is
necessary.

For financial problems, we typically have
50-100 time steps, and about 50 grid points. 
Therefore, a one-dimensional problem is reduced to the computation
of the product of 100 matrices of dimension $50$, which
can be handled on a PC in a few seconds.
The result is usually accurate to a level of $10^{-3}$ without any particular caution.
If a very high precision is required
we can either improve the approximation of the short-time transition
probability or decrease the time step, $\Delta t$.

In the first case, we must expand the short-time
transition probability in powers of $\Delta t$ and $\Delta x$,
by requiring that this expression satisfies up to any given order
the partial derivative (Fokker-Planck) equation associated to the Langevin
equation (\ref{e:1.3})\cite{rct1,rct2b}.
The expansion coefficients are analytical
but rather cumbersome. Each coefficient increases the precision of about one
order of magnitude, but there is a trade-off between the required
precision and the complexity of the analytical expressions.

In the second case, we must either increase the number of grid points
according to the relation between $\Delta t$ and $\Delta x$, or
expand the short-time transition probability over some basis of interpolating
functions\cite{taddei}. In the latter method the relation
${\overline\sigma}^2\,\Delta t\sim \Delta x^2$ is not essential anymore.
Obviously we have an additional cost, in terms of CPU time requirements,
for the computation of the expansion coefficients.
Again there is a trade-off between the required precision and the complexity
of the method.

\subsection{Advantages and limits}

The main advantages of the GFDNM with respect to the
Monte Carlo methods are:
\begin{itemize}
\item[-] higher velocity and accuracy;

\item[-] the path dependent derivative securities are easily handled;

\item[-] the solution for all initial values of the stochastic variable
is obtained in a single iteration.
\end{itemize}
The drawback of this approach appears in the multi-dimensional
case. In a $d$-dimensional problem, in general,
we need matrices of dimension $M^d$.
For example, in four dimensions with 50 grid points,
we have matrices of dimension $50^4 = 6250000$. Although such matrices are
sparse (as a result of the Gaussian form of the transition probability),
i.e. a large number of their elements are zero,
they must be stored on the hard disk, increasing the computing time.

However, we do not need the transition probability matrix itself,
but just to multiply this matrix with a vector of $M^d$ points, which,
at least in four dimension, can be easily stored in the RAM of a PC.
On the other hand, the short-time Green function is an analytical function
which can be computed wasting some CPU time,
but without troubles of memory requirement.

Obviously, as $d$ increases, the dimension of the vector
$M^d$ grows up, and the problem becomes quickly intractable. 
In this case, the only numerical method available
is the Monte Carlo, which always gives an answer
(but not necessarily correct).
The comment in bracket is not sarcastic:
in high dimensional problems the random
sampling allows to get a definite answer, but the
configurations taken into account are only a very small part of the possible
ones. Thus, the probability of missing some critical set of
paths is not negligible.

\section{Numerical results}
\label{s:5}

In the following we apply the Green function deterministic numerical method
to some specific problems.

First, in order to compare the numerical
results to the exact ones, we consider three examples with analytical
solutions. In particular, we compute the price of a European
option on a non-dividend-paying stock in the Black and Scholes
model\cite{blasch}, and the prices of
a zero-coupon bond, and a caplet in the Vasicek model\cite{vasicek}.

Second, we apply the method to two path dependent derivatives. In particular,
we compute the prices of an American option on a non-dividend-paying stock
in the Black and Scholes
model, and of a Bermudan swaption in the Vasicek model.

We want to stress that the choice of the above financial
models is only due to the fact that we can compare the numerical results to the
exact solutions. The numerical code, of course, is very general, and
does not depend on this choice. The program also assume a possible time
dependence of the parameters, and could be optimized
for constant parameters saving much CPU time.

\subsection{European option}
\label{s:5.1}

A European option gives the holder the right (and not the obligation)
to {\em buy}\/ or to {\em sell}\/
an underlying asset on
a certain date ({\em exercise date}, or {\em maturity}\/)
for a certain price ({\em exercise price}).
In the first case it is called {\em call
option}\/, in the second one {\em put option}\/.
If $S(T)$ is the price of the underlying asset at the maturity, $T$,
and $\chi$ is the exercise price, the expected value of a European put
option at the time, $t$, in a risk-neutral world is
\begin{equation}
O_{\rm E}\,(s_t,t,T)=E\left[e^{-r\,(T-t)}\;\max(\chi-S(T),0)\mid S(t)=s_t
\,\right],
\label{e:pute}
\end{equation}
where the risk-free rate of interest, $r$, is assumed constant for the
whole life of the option. 
The functional, $g[s(\tau)]$, is then given by
\begin{equation}
g[s(\tau)]=e^{-r\,(T-t)}\;\max(\chi-s(T),0).
\end{equation}
The expression above can be discretized simply as
\begin{equation}
g[s(\tau)]\simeq \max(\chi-s_N,0)\;\prod_{i=0}^{N-1}e^{-r\,\Delta t}\;,
\end{equation}
and the linear array, $g^{(N)}_\alpha$, is
\begin{equation} 
g^{(N)}_\alpha=\max(\chi-s_\alpha,0).
\end{equation}

Furthermore, since the coefficients of the Lagrangian for the Black and Scholes
model are $\sigma(S)=\sigma\, S$, and $A(S)=\mu\, S$,
the matrix, $G_{\alpha\beta}$, is given by
\begin{equation}
G_{\alpha\beta}=w_\alpha\;{1\over\sqrt{2\pi\,\sigma^2\,s_\beta^2\,\Delta t}}\;
\exp\left\{-{(s_\alpha-s_\beta-\mu\,s_\beta\;\Delta t)^2
\over 2\;\sigma^2\,s_\beta^2\;\Delta t}-r\,\Delta t\right\},
\label{e:gblasch}
\end{equation}
where $s_\alpha$ are the grid points, $w_\alpha$ are the weights of the
integration rule, and $\alpha,\beta=1,\ldots,M$.
In Table \ref{t:1} we show a comparison between analytical and
numerical results.

\subsection{Zero-coupon bond}
\label{s:zeroc}

A zero-coupon bond is a contract which yields a certain amount
({\em principal}\/) on a certain date ({\em maturity}\/) in the future. If,
for the sake of simplicity, the principal is equal 1, the price at the time,
$t$, is given by the conditional expectation value of the functional
(see, paper I)
\begin{equation}
g[r(\tau)]=e^{-\int_t^Tr(\tau)\,d\tau},
\label{e:zeroc}
\end{equation}
where $T$ is the maturity. The functional above can be discretized by the
pre-point rule as
\begin{equation}
g[r(\tau)]\simeq\prod_{i=0}^{N-1} e^{-r_i\,\Delta t},
\end{equation}
with $\displaystyle{\Delta t={T-t\over N}}$.
Hence the linear array, $g^{(N)}_\alpha$, is
\begin{equation} 
g^{(N)}_\alpha=\left(
\begin{array}{c} 
1\\
1\\
.\\
.\\
.\\
1\\
\end{array} \right)
\end{equation}
The coefficients of the Lagrangian (\ref{e:lagra})
are $\sigma(r)=\sigma$,
and $A(r)=a(b-r)$.
Therefore the matrix, $G_{\alpha\beta}$, is given by
\begin{equation}
G_{\alpha\beta}=w_\alpha\;{1\over\sqrt{2\pi\,\sigma^2\,\Delta t}}\;
\exp\left\{-{(z_\alpha-z_\beta-a(b-z_\beta)\;\Delta t)^2
\over 2\;\sigma^2\;\Delta t}-z_\beta\,\Delta t\right\},
\label{e:gvas}
\end{equation}
where $z_\alpha$ are the grid points.
Now, by performing the product (\ref{e:prodg}), we obtain directly
the zero-coupon bond price, $P(z_\alpha,t,T)$, for each initial short term
interest rate, $z_\alpha$.

In Table \ref{t:2} we show a comparison between analytical and
numerical results.

\subsection{Caplet}

A more complex example is the computation of the caplet price.
An interest rate cap is a contract which guarantees that the rate charged on
a loan does not exceed a specified value, the {\em cap rate}\/. The cap
can be viewed as a portfolio of European put options on zero-coupon bonds.
The individual options comprising a cap are referred to as caplets.
The expected value of a caplet is given by the expectation value of the
functional (see also paper I),
\begin{eqnarray}
g[r(\tau)]
&=&e^{-\int_t^Tr(\tau)d\tau}\; (\chi-P(r_T,T,s))\;
\theta\Bigl(\chi-P(r_T,T,s)\Bigr)\\
&=&e^{-\int_t^Tr(\tau)d\tau}\;\max(\chi-P(r_T,T,s),0).
\end{eqnarray}
In order to calculate the value of the caplet, we first
need the price of the zero-coupon bond at the time, $T$, which
can be computed by the method described above.
This procedure (unlike a Monte Carlo approach) gives directly the
price for any grid point, $z_\alpha$, and allows direct
integration over this variable. Once the zero-coupon bond price has been
computed, the vector, $g^{(N)}_\alpha$, is given by
\begin{equation} 
g^{(N)}_\alpha=\max(\chi-P(z_\alpha,T,s),0),
\end{equation}
while the matrix, $G_{\alpha\beta}$, is still given by
the expression (\ref{e:gvas}).

\begin{figure}
\centering
\includegraphics[width=12cm,bb=0 140 612 600,clip]{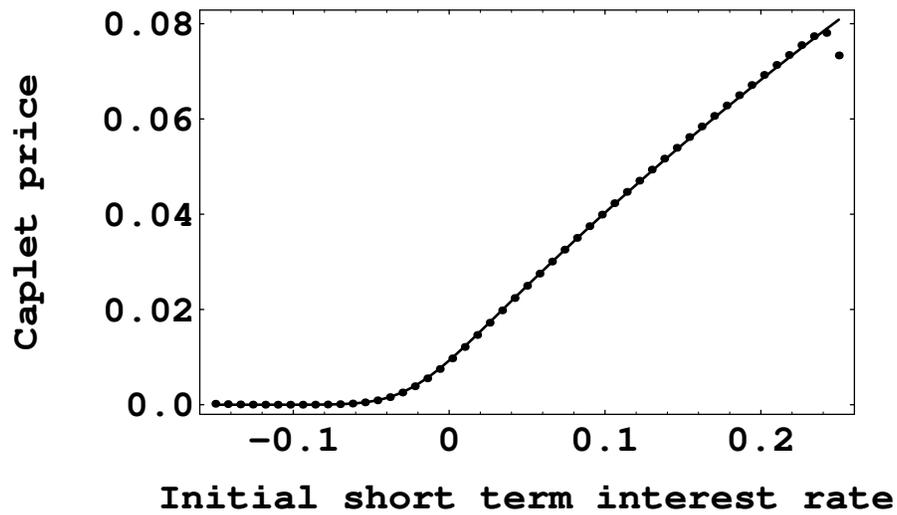}
\caption{Comparison of the analytical Vasicek solution (solid line) with
the numerical solution (dots) obtained by the Green function
deterministic numerical method for a caplet with $\chi=1.001$,
$t=0$ yrs, $T=0.5$ yrs, and $s=1$ yrs.
The Vasicek parameters are
$a=0.5$, $b=0.05$, and $\sigma=0.03$. The number of grid points is 51,
the relative error is of the order of $10^{-2}$, and
the CPU time is much less than 1 second on a Pentium 133.
\label{f:1}}
\end{figure}

In Fig. \ref{f:1} we show a comparison between analytical and
numerical results.

\subsection{American option}
\label{s:5.4}

A path dependent option is an option whose value depends on the past history
of the underlying asset, not just on its value on exercise.
As a first example of a path dependent option, we consider an
American put option, i.e. an options which can be exercised at any time up to
the expiration date. Unfortunately an exact analytical expression for
the price of an American option does not exist. Therefore the importance
of a powerful numerical technique which is able to handle
this problem is evident. The Monte Carlo method is
not very appropriate in this case, since has the disadvantage that
early exercise features are difficult to implement.
The techniques usually adopted are based
on the binomial trees\cite{hull} or on the resolution of
partial differential equations by finite differences\cite{will}.
The Green function deterministic numerical method described in the previous
sections is particularly efficient, in terms of convergence properties,
memory requirements, accuracy, and implementation.

In section \ref{s:5.1} we have seen that, if $S(T)$ is the price of the
underlying asset at the
maturity, $T$, and $\chi$ is the exercise price, the expected value of a
European put option at the time, $t$, in a risk-neutral world is
given by Eq. (\ref{e:pute}). On the other hand, if the option is American,
its value cannot be defined so easily.
From a computational point of view, however, it can be obtained as follows.
The option price, in general, is evaluated by starting at the final time, $T$,
where the values of the European and the American options coincide,
and are simply given by
\begin{equation}
O_{\rm A}\,(S(T),T,T)=O_{\rm E}\,(S(T),T,T)=\max(\chi-S(T),0),
\end{equation}
and working backward in time steps, $\Delta t$.
If the option is American, we also need to check at
any time, and for any value of the stock price, whether early
exercise is preferable to holding the option for a further time step.
Therefore the procedure for the calculation is the following:
\begin{itemize}
\item[i.] We compute the value of the option at the time, $T-\Delta t$,
if it is not exercised:
\begin{equation}
E\left[e^{-r\,\Delta t}\;O_{\rm A}\,(S(T),T,T)\mid S(T-\Delta t)=
s_{T-\Delta t}\,\right]
\end{equation}
(we recall that the risk-free rate of interest, $r$, is constant here).

\item[ii.] We compute the value of the option at the time, $T-\Delta t$,
if it is exercised:
\begin{equation}
\max(\chi-s_{T-\Delta t},0).
\end{equation}

\item[iii.] The correct expected value of the option at the time, $T-\Delta t$,
is given by
\begin{eqnarray}
\lefteqn{\!\!\!\!\!\!\!\!\!\!\!\!O_{\rm A}\,(s_{T-\Delta t},T-\Delta t,T)=}
\nonumber\\
&&\max\Bigl(
E\left[e^{-r\,\Delta t}\;
O_{\rm A}\,(S(T),T,T)\mid S(T-\Delta t)=s_{T-\Delta t}\,\right]\;,
\nonumber\\
&&\;\;\;\;\;\;\;\;\;
\max(\chi-s_{T-\Delta t},0)\Bigl),
\end{eqnarray}
i.e. we check if it is convenient to exercise or not the option.

\item[iv.] We put $T= T-\Delta t$, and we iterate the procedure until the
initial time.
\end{itemize}
In conclusion, by recalling the algorithm and the notations given in section
\ref{s:4pathd}, we have
\begin{eqnarray} 
g^{(N)}_\alpha&=&\max(\chi-s_\alpha,0),\\
w^{(i)}_\alpha&=&\max(\chi-s_\alpha,0),
\end{eqnarray}
and the matrix, $G_{\alpha\beta}$, is the same of Eq. (\ref{e:gblasch}).

In Table \ref{t:3} we show a comparison of the numerical results
obtained by a finite-difference method\cite{will},
a binomial method\cite{hull}, and the Green function deterministic numerical
method.

\subsection{Bermudan swaption}
\label{s:5.5}

A swap contract is an agreement between two companies to exchange cash
flows in the future according to a defined formula.
The simplest example of swap is the {\em plain vanilla} interest rate swap.
In this case a first party agrees to pay to a second party cash flows equal to
interest at a predeterminate fixed rate on a principal for a
number of years. At the same time the second party agrees to pay to the
first party cash flows equal to
interest at a floating rate on the same principal for the same
period of time, in the same currency. Actually, there is an enormous number
of different swap types that can be invented. For the sake of
simplicity, we will consider only the plain vanilla. However, the
method described in this section can be easily extended to more complex cases.

A swaption is an option on a swap contract.
It gives the holder the right to enter into or to terminate a swap
contract at a certain time in the future.
In the following we will consider a Bermudan swaption, which is a particular
non-standard American option. In a Bermudan swaption early exercise is
restricted to certain dates during the life of the swap, usually the
reset dates.

Let us then consider a plain vanilla interest rate swap settled in arrears,
with a principal, $Q$, a fixed rate, $K$,
the floating rate equal to the London Interbank Offer Rate (LIBOR),
the fixed rate payment dates, $t_1^K,\ldots,t_{N_K}^K$,
and the floating rate payment dates, $t_1^L,\ldots,t_{N_L}^L$. We can assume
that the floating base rate underlying the swap is the appropriate rate to use
for discounting. This is a common assumption (see, for example, section 5.2
in Ref.\cite{hull}), and considerably simplifies the valuation procedure.
Therefore the floating rate, $L_i$, at the time, $t_i^L$, is given by
\begin{equation}
L_i\equiv L(R(t_i^L),t_i^L,t_{i+1}^L)={1\over\Delta_L}\;
\left({1\over P(R(t_i^L),t_i^L,t_{i+1}^L)}-1\right).
\label{e:libor}
\end{equation}
The swap price, $W({\overline t})$, at the time, ${\overline t}$,
is given by the sum of the
actualized values of all remaining payments after ${\overline t}$.
If we define the integers, ${\overline N}_K$, ${\overline N}_L$, with
$1\leq {\overline N}_K\leq N_K$, and $1\leq {\overline N}_L\leq N_L$, such that
$t_{{\overline N}_K}^K$, and $t_{{\overline N}_L}^L$ are the dates of the first
fixed rate and floating rate payments after ${\overline t}$, respectively,
we have
\begin{equation}
W({\overline t})=\sum_{i={{\overline N}_L}}^{N_L}
W_i^L({\overline t},t_i^L)-
\sum_{i={{\overline N}_K}}^{N_K} W_i^K({\overline t},t_i^K),
\end{equation}
where
\begin{eqnarray}
W_i^L({\overline t},t_i^L)&=&E\left[
e^{-\int_{\overline t}^{t_i^L}R(\tau)\,d\tau}\;Q\;
L_{i-1}\,\Delta_L\mid
R(\tau)=r_\tau,\,\forall\;\tau\leq{\overline t}\,\right],\\
W_i^K({\overline t},t_i^K)&=&E\left[
e^{-\int_{\overline t}^{t_i^K}R(\tau)\,d\tau}\;Q\;K\;\Delta_K\mid
R(\tau)=r_\tau,\,\forall\;\tau\leq{\overline t}\,\right],
\end{eqnarray}
$\Delta_K=t_i^K-t_{i-1}^K$, $\Delta_L=t_i^L-t_{i-1}^L$, and
$t_0^K=t_1^K-\Delta_K$, $t_0^L=t_1^L-\Delta_L$. Note that the
floating rate payment at the time $t_i^L$ is made according the rate at the
beginning of the period, i.e. in arrears.
Since $R(\tau)$ is a Markov process, the expressions above are equivalent to
\begin{eqnarray}
&&\!\!\!\!\!\!\!\!\!\!W_i^L({\overline t},t_i^L)\!=\!\left\{
\begin{array}{ll}
\!\!\!E\!\left[
e^{-\int_{\overline t}^{t_i^L}\!\!R(\tau)\,d\tau}Q
L_{i-1}\Delta_L\!\mid\!
R({\overline t})\!=r_{\overline t}\right]\!,&
\!\!\!{\overline t}\!\leq\! t_{i-1}^L\\
\\
\!\!\!E\!\left[
e^{-\int_{\overline t}^{t_i^L}\!\!R(\tau)\,d\tau}Q
L_{i-1}\Delta_L\!\mid\!
R({\overline t})\!=r_{\overline t}\,;R(t_{i-1})\!=r_{t_{i-1}}\!\right]\!,&
\!\!\!t_{i-1}^L\!\!<\!{\overline t}\!\leq\! t_i^L
\end{array} \right.
\label{e:sl}\\
\nonumber\\
\nonumber\\
&&\!\!\!\!\!\!\!\!\!\!W_i^K({\overline t},t_i^K)=E\!\left[
e^{-\int_{\overline t}^{t_i^K}\!\!R(\tau)\,d\tau}\,Q\,K\,\Delta_K\!\mid\!
R({\overline t})\!=r_{\overline t}\,\right].
\label{e:sr}
\end{eqnarray}
By exploiting the expression (\ref{e:libor}) for the LIBOR,
and the properties (\ref{e:innerpath}), and (\ref{e:finalpath}), the Eqs.
(\ref{e:sl}), and (\ref{e:sr}), become
\begin{eqnarray}
\lefteqn{\!\!W_i^L({\overline t},t_i^L)=}\nonumber\\
\nonumber\\
&&\!\!\!\!\!\!\!\!\!\!\!\!\!\!\!\!
\left\{
\begin{array}{ll}
Q\,\Biggl(E\Biggl[\;\,
E\Biggl[e^{-\int_{t_{i-1}^L}^{t_i^L}\!R(\tau)\,d\tau}\mid
R(t_{i-1}^L)\Biggr]\;
\displaystyle{e^{-\int_{\overline t}^{t_{i-1}^L}R(\tau)\,d\tau}
\over P(R(t_{i-1}^L),t_{i-1}^L,t_i^L)}\mid
R({\overline t})=r_{\overline t}\;\Biggr]&\\
\\
\;\;\;\;\;\;\;\;\;\;\;\;\;\;\;\;\;\;\;\;\;\;\;\;
-\;\;E\left[e^{-\int_{\overline t}^{t_i^L}R(\tau)\,d\tau}\mid
R({\overline t})=r_{\overline t}\;\right]\Biggr),&
\!\!\!\!\!\!\!\!\!\!\!\!\!\!\!\!\!\!\!\!\!\!
{\overline t}\!\leq\! t_{i-1}^L\\
\\
Q\,\Biggl(P(r_{t_{i-1}^L},t_{i-1}^L,t_i^L)^{-1}\;
E\Biggl[e^{-\int_{\overline t}^{t_i^L}\!R(\tau)\,d\tau}\mid
R({\overline t})=r_{\overline t}\;\Biggr]\\
\\
\;\;\;\;\;\;\;\;\;\;\;\;\;\;\;\;\;\;\;\;\;\;\;\;
-\;\;E\left[e^{-\int_{\overline t}^{t_i^L}R(\tau)\,d\tau}\mid
R({\overline t})=r_{\overline t}\;\right]\Biggr),&
\!\!\!\!\!\!\!\!\!\!\!\!\!\!\!\!\!\!\!\!\!\!
t_{i-1}^L\!\!<\!{\overline t}\!\leq\! t_i^L
\end{array} \right.
\nonumber\\
\nonumber\\
\nonumber\\
&=&\left\{
\begin{array}{ll}
Q\left(
P(r_{\overline t},{\overline t},t_{i-1}^L)
-P(r_{\overline t},{\overline t},t_i^L)\right),&
\;\;\;\;\;\;\;\;\;\;\;\;\;\;\;
{\overline t}\!\leq\! t_{i-1}^L\\
\\
Q\left(
\displaystyle{P(r_{\overline t},{\overline t},t_i^L)\over
P(r_{t_{i-1}^L},t_{i-1}^L,t_i^L)}
-P(r_{\overline t},{\overline t},t_i^L)\right),&
\;\;\;\;\;\;\;\;\;\;\;\;\;\;\;
t_{i-1}^L\!\!<\!{\overline t}\!\leq\! t_i^L
\end{array} \right.\\
\nonumber\\
\nonumber\\
&&\!\!\!\!\!\!\!\!\!\!\!\!\!\!\!\!
W_i^K({\overline t},t_i^K)=Q\;K\;\Delta_K\;
P(r_{\overline t},{\overline t},t_i^K).
\end{eqnarray}
In conclusion, the calculation of the expected value of a swap contract can
be reduced to that of zero-coupon bonds, and can be made
by using the procedure described in section \ref{s:zeroc}.

Let us now consider a Bermudan swaption which gives the right to terminate
the swap contract defined above at various dates, $T_1,\ldots,T_N$.
If we assume that the exercise dates, $T_i$, coincide with some floating rate
payment dates, $t_j^L$, as usual, the swap price at the time,
${\overline t}=T_i$,
depends on the short term interest rate, $r_{\overline t}$,
only, i.e. $W({\overline t})\equiv W(r_{\overline t},{\overline t})$.
The expected value at the time, $t<T_1$, of the corresponding European swaption
is simply
\begin{equation}
O_{\rm E}^W\,(r_t,t,,T_N)=
E\left[e^{-\int_t^{T_N}R(\tau)\,d\tau}
\max\left(W(R(T_N),T_N)\,,\,0\right)\mid R(t)=r_t\,\right].
\end{equation}
In the case of a Bermudan swaption, we also need to check at all times, $T_i$,
and for any possible value of the short term interest rate, whether early
exercise is preferable to holding the swaption for a further time interval.
The procedure is analogous to that of an American option. The main difference
is that the underlying asset price is not the stochastic variable itself, but
the swap price. Therefore the swap price must be calculated in advance for
all exercise dates of the swaption.
We use an algorithm similar to that of section \ref{s:4pathd} with
\begin{eqnarray} 
g^{(N)}_\alpha&=&\max\left(W(r_\alpha,T_N)\,,\,0\right),\\
w^{(i)}_\alpha&=&\max\left(W(r_\alpha,T_i)\,,\,0\right),
\end{eqnarray}
and the matrix, $G_{\alpha\beta}$, given in Eq. (\ref{e:gvas}).
The only difference is that the test in step ii is made only at the
exercise dates of the swaption.

In Table \ref{t:4}, we show the numerical results obtained for a
plain vanilla interest rate swap in the Vasicek model
compared with the analytical ones. In Table \ref{t:5}, and
Table \ref{t:6} we show a comparison of the numerical solutions
obtained for the European and the Bermudan swaptions, respectively,
with the results obtained by a semi-analytical computation.

\subsection{Greeks}

Any financial institution has the problem of hedging the risk of its
portfolio of derivative securities. For this reason it needs to know the
sensitivity of the portfolio to the changes of the underlying asset prices,
the time, and the market conditions. This sensitivity is usually measured by
calculating five hedge parameters (greeks):
\begin{itemize}
\item {\em delta},\/ is the rate of change of the derivative security
price with respect to the price of the underlying asset;
\item {\em gamma},\/ is the rate
of change of the portfolio's delta with respect to the price of the underlying
asset;
\item {\em theta},\/ is the rate of change of the value of the portfolio
with respect to time;
\item {\em vega},\/ is the rate of change of the value of the
portfolio with respect to the volatility of the underlying asset;
\item {\em rho},\/
is the rate of change of the value of the portfolio with respect to the
interest rate.
\end{itemize}
The Green function deterministic method gives directly the price of a
derivative security as a discretized function of the initial value of the
underlying variable, and the whole discrete time evolution of the price
is calculated step by step. Moreover, the price variation with respect to
other parameters can be obtained by changing these parameters, and
performing a new computation. Therefore, the calculation of the
greeks does not add any further complication.

\subsection{Conclusions}

We have described two numerical methods, based on the path integral formulation
given in Ref. \cite{rct1}, to calculate the conditional expectation value of a
general functional: the Monte Carlo method,
and the Green function deterministic
numerical method. Moreover,
we have shown some practical applications of the latter
to the pricing of derivative securities. In order to compare the analytical and
the numerical results, we used two solvable financial models:
the Black-Scholes, and the Vasicek models. However, the GFDNM
works also in more complex cases, when an analytical solution does not
exist. The numerical results are very
accurate, and can be even improved by using some particular
techniques\cite{rct1,rct2b,taddei}.
The method has been applied to one-dimensional problems, but the extension
to the $d$-dimensional ($d\sim 2- 4$) ones is straightforward.
Another important feature of the GFDNM is that it gives directly the
conditional expectation value of a functional for all
initial values of the stochastic variable.
Finally, the case of the path dependent derivative securities
can be handled with only small changes in the code.

In conclusion, the GFDNM is a very powerful technique for low dimensional
problems, as usually the derivative security pricing. On the other hand,
the Monte Carlo method is the only
possible, although very slow and imprecise, for high dimensional problems.

\clearpage

\renewcommand{\footnoterule}{\vskip -8pt}

\begin{table}
\centering
\begin{minipage}{\textwidth}
\caption{Comparison of the analytical Black-Scholes solution with
the numerical solutions
obtained by a binomial method, and the Green function
deterministic numerical
method for a European put option with $\chi=10$,
$t=0$ yrs, and $T=0.5$ yrs.}
\footnotetext{{\hskip -8pt}{\em Note.\/} The Black-Scholes parameters are
$r=0.1$, and $\sigma=0.4$. The number of grid points for the GFDNM is 201,
and the CPU time is about 3 seconds on a Pentium 133.}
\label{t:1}
\begin{tabular*}{\textwidth}{@{\extracolsep{\fill}}cccc}
\hline\hline
Stock price&Analytical&Binomial\cite{hull}&GFDNM\\
\hline
6.0&3.558&3.557&3.557\\
8.0&1.918&1.917&1.917\\
10.0&0.870&0.866&0.871\\
12.0&0.348&0.351&0.349\\
14.0&0.128&0.128&0.129\\
\hline\hline
\end{tabular*}
\end{minipage}
\end{table}

\begin{table}
\centering
\begin{minipage}{\textwidth}
\caption{Comparison of the analytical Vasicek solution with
the numerical solution obtained by the Green function
deterministic numerical method for a zero-coupon bond with 
$t=0$ yrs, and $T=0.5$ yrs.}
\footnotetext{{\hskip -8pt}{\em Note.\/} The Vasicek parameters are
$a=0.5$, $b=0.05$, and $\sigma=0.03$. The number of grid points is 51, and
the CPU time is much less than 1 second on a Pentium 133.}
\label{t:2}
\begin{tabular*}{\textwidth}{@{\extracolsep{\fill}}cccc}
\hline\hline
Short term&&&\\
interest rate&Analytical&GFDNM&Relative error\\
\hline
0.02&0.9884&0.9886&$2\times 10^{-4}$\\
0.04&0.9797&0.9797&$8\times 10^{-5}$\\
0.06&0.9710&0.9709&$8\times 10^{-5}$\\
0.08&0.9625&0.9622&$3\times 10^{-4}$\\
\hline\hline
\end{tabular*}
\end{minipage}
\end{table}

\begin{table}
\centering
\begin{minipage}{\textwidth}
\caption{Comparison of the numerical Black-Scholes solutions
obtained by a finite-difference method,
a binomial method, and the Green function deterministic numerical
method for an American put option with $\chi=10$,
$t=0$ yrs, and $T=0.5$ yrs.}
\footnotetext{{\hskip -8pt}{\em Note.\/} The Black-Scholes parameters are
$r=0.1$, and $\sigma=0.4$. The number of grid points for the GFDNM is 201,
and the CPU time is about 3 seconds on a Pentium 133.}
\label{t:3}
\begin{tabular*}{\textwidth}{@{\extracolsep{\fill}}cccc}
\hline\hline
Stock price&Finite difference\cite{will}&Binomial\cite{hull}&GFDNM\\
\hline
6.0&4.000&4.000&4.000\\
8.0&2.095&2.096&2.093\\
10.0&0.921&0.920&0.922\\
12.0&0.362&0.365&0.364\\
14.0&0.132&0.133&0.133\\
\hline\hline
\end{tabular*}
\end{minipage}
\end{table}

\begin{table}
\centering
\begin{minipage}{\textwidth}
\caption{Comparison of the analytical Vasicek
solution with
the numerical solution obtained by the Green function
deterministic numerical method for a swap with $Q=1$, $K=0.045$,
${\overline t}=0$ yrs,
$t_0^K=t_0^L=10$ yrs, $\Delta_K=\Delta_L=0.5$ yrs, and $N_K=N_L=10$.}
\footnotetext{{\hskip -8pt}{\em Note.\/} The Vasicek parameters are
$a=0.5$, $b=0.05$, and $\sigma=0.03$. The number of grid points is 51, and
the CPU time is about 3 seconds on a Pentium 133.}
\label{t:4}
\begin{tabular*}{\textwidth}{@{\extracolsep{\fill}}cccc}
\hline\hline
Short term&&&\\
interest rate&Analytical&GFDNM&Relative error\\
\hline
0.02&0.01064&0.01066&$2\times 10^{-3}$\\
0.04&0.01037&0.01037&$4\times 10^{-4}$\\
0.06&0.01010&0.01009&$9\times 10^{-4}$\\
0.08&0.00984&0.00982&$2\times 10^{-3}$\\
\hline\hline
\end{tabular*}
\end{minipage}
\end{table}

\begin{table}
\centering
\begin{minipage}{\textwidth}
\caption{Comparison of a semi-analytical Vasicek
solution with
the numerical solution obtained by the Green function
deterministic numerical method for a European swaption with
$Q=1$, $K=0.045$, $t=0$ yrs,
$t_0^K=t_0^L=10$ yrs, $\Delta_K=\Delta_L=0.5$ yrs, and $N_K=N_L=10$,
which gives the right to terminate the swap
just after the $8^{\rm th}$ payment date.}
\footnotetext{{\hskip -8pt}{\em Note.\/} The Vasicek parameters are
$a=0.5$, $b=0.05$, and $\sigma=0.03$.
The CPU time with 51 grid points is about 3 seconds on a Pentium 133.}
\label{t:5}
\begin{tabular*}{\textwidth}{@{\extracolsep{\fill}}cccccc}
\hline\hline
&Semi-analytical&\multicolumn{2}{c} {GFDNM (51 pts)}&
\multicolumn{2}{c} {GFDNM (101 pts)}\\
&&&&&\\
Short term&Swaption&Swaption&Relative&Swaption&Relative\\
interest rate&price&price&error&price&error\\
\hline
0.02&0.00409&0.00423&$3\times 10^{-2}$&0.00412&$7\times 10^{-3}$\\
0.04&0.00393&0.00405&$3\times 10^{-2}$&0.00396&$8\times 10^{-3}$\\
0.06&0.00377&0.00389&$3\times 10^{-2}$&0.00380&$8\times 10^{-3}$\\
0.08&0.00362&0.00372&$3\times 10^{-2}$&0.00365&$8\times 10^{-3}$\\
\hline\hline
\end{tabular*}
\end{minipage}
\end{table}

\begin{table}
\centering
\begin{minipage}{\textwidth}
\caption{Comparison of a semi-analytical Vasicek
solution with
the numerical solution obtained by the Green function
deterministic numerical method for a Bermudan swaption with
$Q=1$, $K=0.045$, $t=0$ yrs,
$t_0^K=t_0^L=10$ yrs, $\Delta_K=\Delta_L=0.5$ yrs, and $N_K=N_L=10$,
which gives the right to terminate the swap at the time $t_0^K$, and
just after the
$2^{\rm nd},4^{\rm th},6^{\rm th}$, and $8^{\rm th}$ payment dates.}
\footnotetext{{\hskip -8pt}{\em Note.\/} The Vasicek parameters are
$a=0.5$, $b=0.05$, and $\sigma=0.03$.
The CPU time with 51 grid points is about 3 seconds on a Pentium 133.}
\label{t:6}
\begin{tabular*}{\textwidth}{@{\extracolsep{\fill}}cccccc}
\hline\hline
&Semi-analytical&\multicolumn{2}{c} {GFDNM (51 pts)}&
\multicolumn{2}{c} {GFDNM (101 pts)}\\
&&&&&\\
Short term&Swaption&Swaption&Relative&Swaption&Relative\\
interest rate&price&price&error&price&error\\
\hline
0.02&0.01559&0.01609&$3\times 10^{-2}$&0.01569&$6\times 10^{-3}$\\
0.04&0.01494&0.01539&$3\times 10^{-2}$&0.01503&$6\times 10^{-3}$\\
0.06&0.01431&0.01472&$3\times 10^{-2}$&0.01440&$6\times 10^{-3}$\\
0.08&0.01371&0.01408&$3\times 10^{-2}$&0.01380&$7\times 10^{-3}$\\
\hline\hline
\end{tabular*}
\end{minipage}
\end{table}

\clearpage

\end{document}